# Multi-Contrast Computed Tomography Healthy Kidney Atlas

Ho Hin Lee*†, Yucheng Tang, Kaiwen Xu, Shunxing Bao, Agnes B. Fogo, Raymond Harris,
Mark P. de Caestecker, Mattias Heinrich, Jeffrey M. Spraggins, Yuankai Huo, Bennett A. Landman

*Abstract*— The construction of three-dimensional multi-modal tissue maps provides an opportunity to spur interdisciplinary innovations across temporal and spatial scales through information integration. While the preponderance of effort is allocated to the cellular level and explore the changes in cell interactions and organizations, contextualizing findings within organs and systems is essential to visualize and interpret higher resolution linkage across scales. There is a substantial normal variation of kidney morphometry and appearance across body size, sex, and imaging protocols in abdominal computed tomography (CT). A volumetric atlas framework is needed to integrate and visualize the variability across scales. However, there is no abdominal and retroperitoneal organs atlas framework for multi-contrast CT. Hence, we proposed a high-resolution CT retroperitoneal atlas specifically optimized for the kidney across non-contrast CT and early arterial, late arterial, venous and delayed contrast enhanced CT. Briefly, we introduce a deep learning-based volume of interest extraction method and an automated two-stage hierarchal registration pipeline to register abdominal volumes to a high-resolution CT atlas template. To generate and evaluate the atlas, multi-contrast modality CT scans of 500 subjects (without reported history of renal disease, age: 15-50 years, 250 males & 250 females) were processed. We demonstrate a stable generalizability of the atlas template for integrating the normal kidney variation from small to large, across contrast modalities and populations with great variability of demographics. The linkage of atlas and demographics provided a better understanding of the variation of kidney anatomy across populations.

*Index Terms*—Retroperitoneal atlas, deep learning, multi-contrast computed tomography, medical image registration, kidney atlas

This work is supported by a research fund HuBMAP from NIH.

Ho Hin Lee, Yucheng Tang, Xu Kaiwen, Shunxing Bao, Yuankai Huo and Bennett A. Landman are with the Department of Electrical Engineering and Computer Science, Vanderbilt University, Tennessee, United States.

Jeffery M. Spraggins is with the Department of Biochemistry, Vanderbilt University, Tennessee, United States.

Agnes B. Fogo is with the Department of Pathology, Microbiology and Immunology, Vanderbilt University Medical Center, Tennessee, United States.

Raymond Harris and Mark P. de Caestecker are with the division of Nephrology and Hypertension in Department of Medicine, Vanderbilt University Medical Center, Tennessee, United States.

Mattias Heinrich is with the Institute of Medical Informatics, University of Lübeck, Germany.

*First authors, †Corresponding author, Email: ho.hin.lee@vanderbilt.edu

## I. INTRODUCTION

Physiological and metabolic processes are performed in parallel in the human body. Complicated relationships between cells are required for proper organ function but are challenging to analyze. Extensive studies mapping the organization and molecular profiles of cells within tissues or organs are needed across the human body. While the majority of efforts are distributed to the cellular and molecular perspectives [1], generalizing information from cell to organ level is essential to provide a better understanding of the functionality and linkage across scales [2]. The use of computed tomography provides an opportunity to contextualize the anatomical characteristics of organs and systems in the human body. In addition, by creating a generalizable framework with integration of micro-scale information and system-scale information, this will provide clinicians and researchers the ability to visualize the complex organization of tissues from a cellular to tissue level.

Abdominal CT provides information about abdominal organs at a system scale. Contrast enhancement demonstrates additional anatomical and structural details of organs and neighboring vessels by injecting a contrast agent before imaging procedures. Five different contrast phases are typically generated corresponding to the timing of the contrast agent in the imaging cycle: 1) non-contrast, 2) early arterial, 3) late arterial, 4) portal venous and 5) delayed. The intensity range of organs fluctuates across the contrast enhanced imaging cycle and the variation of intensity helps to capture and specify contextual features of each specific organ. The kidneys, which are located retroperitoneally, also have challenges in imaging. From the anatomical information provided from the contrast enhanced CT of large clinical cohorts, healthy kidney morphometry and appearance may vary. An atlas reference framework is needed to generalize the anatomical and contextual features across the variations in sex, body size, and imaging protocols. However, due to the large variability in anatomy and morphology of various abdominal and retroperitoneal organs, generating a standard reference template for each of these organs is still challenging and no atlas framework for abdominal or retroperitoneal organs is currently publicly available.

Creating an atlas for particular anatomical regions has widely been used with magnetic resonance imaging (MRI). Extensive efforts are allocated in multiple perspectives of brain atlas with brain MRI. *Kuklisova-Murgasova et al.* generated multiple atlases for early developing babies with age ranging from 29 to 44 weeks using affine registration [3], while *Shi et al.* proposed an infant brain atlas using unbiased group-wise registration with three varying scanning time points of brain MRI from 56 males and 39 females normal infants [4]. Unbiased spatiotemporal 4-dimensional MRI atlas and time-variable longitudinal MRI atlas for infant brain were also proposed by *Ali et al.* with diffeomorphic deformable registration with kernel regression in age, and by Yuyao *et al.* patch-based registration in spatial-temporal wavelet domain respectively [5]. Apart from generating atlas framework for normal brain, *Rajashekar et al.* generated two high-resolution normative disease-related brain atlases in FLAIR MRI and non-contrast CT modalities, to investigate lesion-related diseases such as stroke and multiple sclerosis in elderly population [6]. Meanwhile, limited studies have proposed creating a standard reference framework for abdominal and retroperitoneal organs. Development of abdominal and retroperitoneal organ atlases is challenging across multi-modality images (CT and MRI) due to the limited robustness of the registration methods and significant morphology variations in multiple organs associated with patients' demographics [7].

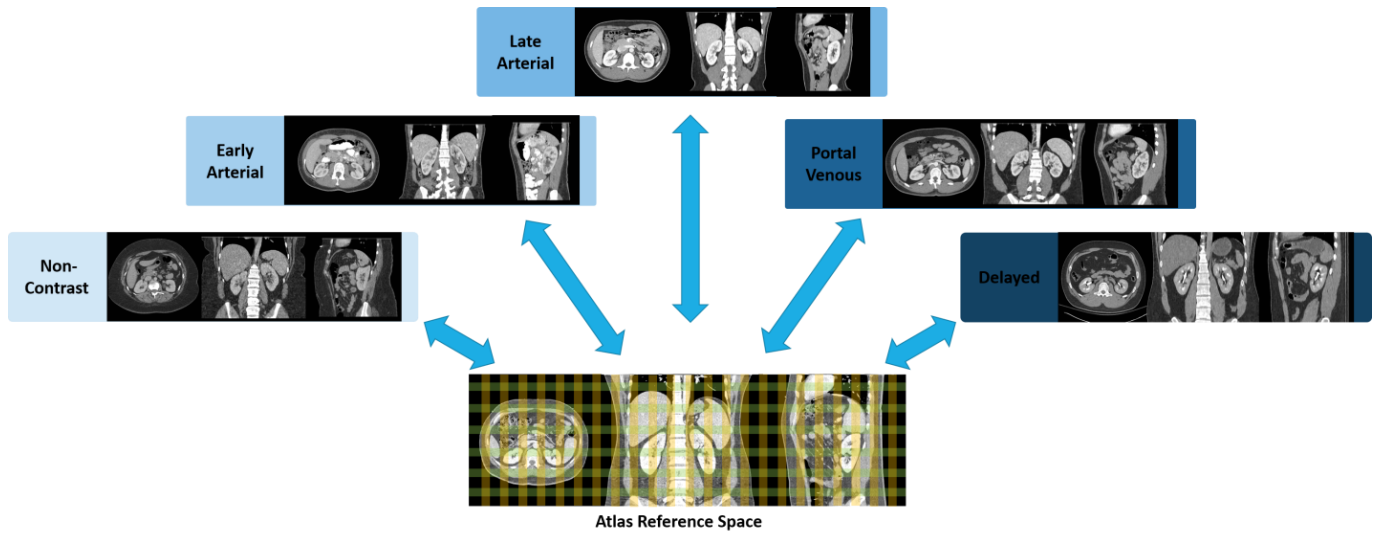

Fig. 1. Illustration of multi-contrast phase CT atlas. The color grid in the three-dimensional atlas space represents the defined spatial reference for the abdominal-to-retroperitoneal volume of interest and localizes abdominal and retroperitoneal organs with each contrast phase characteristics. Blue arrows represent the bi-directional transformation across the atlas target defined spatial reference and the original source image space.

In this work, we present a contrast-preserving CT retroperitoneal atlas framework, optimized for healthy kidney organs with contrast-characterized variability and the generalizable features across a large population of clinical cohorts as shown in Fig. 1. Specifically, as to reduce the failure rate of transferring the morphological and contrastive characteristics of kidney organs, we initially extracted the abdominal-to-retroperitoneal volume of interest with a similar field of view to the atlas target image, using a deep neural network called body part regression (BPR) [8]. 2D slices of the CT volume assessed with BPR model and generate a value ranging from -12 to +12, corresponding to the upper lung region and the pelvis region respectively in the body. By limiting the range of values for both abdominal and retroperitoneal regions, each CT volume is cropped and excludes other regions apart from the abdomen and retroperitoneum, such as the lung and pelvis. A two-stage hierarchical registration pipeline is then performed, registering the extracted volume interest to the high-resolution atlas target with affine and non-rigid registration [9, 10]. To ensure the stability and the variation localized in the atlas template, average and variance mappings across the multi-contrast registered output is computed to demonstrate a better understanding of anatomical details of kidney organs across different contrasts. Overall, our main contributions are summarized as:

- We constructed the first multi-contrast CT healthy kidney atlas framework for the public usage domain.
- We proposed a standardized framework optimizing for kidney organs, and generalized the anatomical context of kidneys with significant variation of morphological and contrastive characteristics across demographics and imaging protocols.
- We evaluate the generalizability of the atlas template by transferring the atlas target label to the 13 organs well-annotated CT space with inverse transformation. An unlabeled multi-contrast phase CT cohort is used to compute average and variance mapping to demonstrate the effectiveness and stability of the proposed atlas framework. Our proposed atlas framework demonstrates a stable transfer ability in both left and right kidneys with median Dice above 0.8.

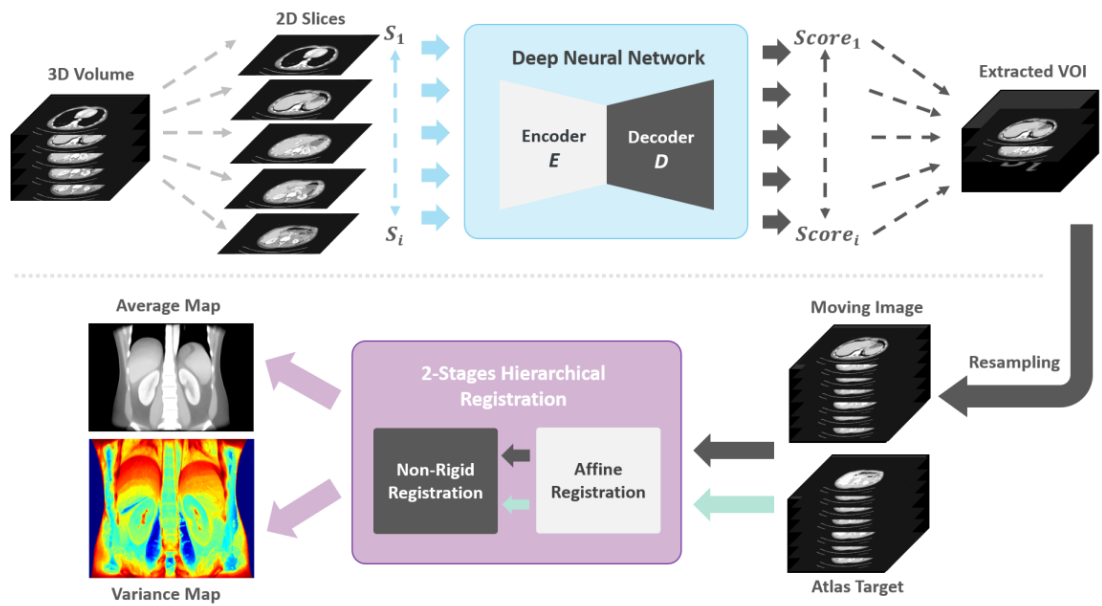

*Figure 2.* The overview of the complete pipeline to create kidney atlas template is illustrated. The input volume is initially cropped to a similar field of view with the atlas target. The extracted volumes of interest are resampled to the same resolution and dimension of the atlas template and performed 2-stages hierarchical registration. The successfully registered scans are finally used to compute the average template and the variance maps.

This work is a significant extension of our prior conference paper [11]. First, we compute the quantitative measures with different registrations methods and evaluate the registration performance of the atlas framework with 13 organs corresponding label transfer using an external portal venous phase labeled cohort. Second, we compute surface renderings with 2D color-space checkerboard pattern to reveal the actual correspondence of the atlas inverse transferred label and the atlas label, visualizing the horizontal and vertical deformation to evaluate the stability of registration. In addition, we provide a more detailed analysis of our proposed atlas framework.

## II. METHODS

Fig. 2 presents an overview of the complete pipeline for generating the kidney atlas framework. The volume of interest is first extracted with a deep learning-based body part regression algorithm to obtain a similar field of view with the atlas target image, increasing the stability of registration in the abdominal body and kidneys. Here, we define the stability as the atlas not changing with randomized subjects and measure the stability with the mean or variance mapping of the atlas template. The integration of deep learning-based volume of interest extraction and classical registration provided an opportunity to reduce the subjectivity of choosing the field of view between source and target image, and increase the [1]robustness of the image registration across the clinical cohorts.

### A. Deep Body Part Regression Network

The use of deep learning in the medical imaging domain contributed to a great increase in automatic models for classification and segmentation. Due to the shift of various domains and the variation of imaging protocols, medical images usually present with different visual appearances and fields of view. The goal of the body part regression network (BPR) is to narrow the difference of field of view between the source images and the atlas target image for reducing the failure rate of registration. Formally, given an unlabeled dataset $\{x_i^m\}_{i=1}^N$ from the moving image domain, and a labeled dataset $\{x_i^a, y_i^a\}_{i=1}^1$ from the atlas target domain, we aim to crop the original volume of interest $x^m$ to an approximate field of view with the atlas target $x^a$. The obtained volume of interest only consists of abdomen regions and is resampled to the same voxel resolution as the atlas target.

Yan *et al.* proposed a self-supervised method to predict a continuous score for each axial slice of CT volumes as the normalized body coordinate value without any labels [8]. The self-supervised model predicts scores in the range of -12 to +12 and each body part region corresponds well to an approximate score (e.g., -12: upper chest, -5: diaphragm / upper liver, 4: lower retroperitoneum, 6: pelvis). Linear regression is performed to correct the discontinuity of the predicted score, and we use the regressed output as the self-supervised label to train a new refined model. Both the atlas target image and the unlabeled dataset are input into the well-trained model and compute scores for each slices of the volume. To extract the abdominal-to-retroperitoneal regions only for each dataset, we limit the slices with a range of scores from -5 to 5 and crop the slices that are out of this range. All unlabeled datasets $x^m$ are then enforced to have a closer field of view to the atlas target image $x^a$.

*B. 2-Stage Hierarchal Registration Pipeline*

Before we performed registrations, each source image (moving/floating) is resampled to the same voxel resolution and the 3-dimensional shape of the atlas target (fixed/target) with cropping/padding slices. Our registration pipeline is composed of 2 hierarchal stages: 1) affine registration and 2) non-rigid registration. Dense displacement sampling (DEEDS) is a 3D medical registration tool with a discretized sampling space that has been shown to yield a great performance in abdominal and retroperitoneal organs registration, is used for both affine non-rigid registration in this pipeline [7, 9, 10]. The DEEDS affine registration is first performed to initially align both moving images and the atlas target to preserve 12 degrees of freedom of transformation and provide a prior definition of the spatial context and each affine component. An affine transformation matrix is generated as the output and become the second stage non-rigid registrations' input. The DEEDS non-rigid registration is refined with the spatial context as the local voxel-wise correspondence with its specific similarity metric, which will be illustrated below. Five different scale levels are used with grid spacing ranging from 8 to 4 voxels to extract patches and displacement search radii from six to two steps between 5 and 1 voxels [7, 9, 10, 12]. Deformed scans with the displacement data from selecting control points is generated and transfer the source image space voxel information to the atlas target space after deformation. To ensure the stability of the atlas generated, all successfully deformed scans are averaged and variance mapping is used to visualize the intensity fluctuation and variation around the abdominal body and kidney organs.

The similarity metric defined in DEEDS registration tool is self-similarity context with the patches extracted from moving images. Such a similarity metric aims to find a similar context around neighboring voxels in patches [12]. The self-similarity metric $S$ is optimizing a distance function $D$ between the image patches extracted from the moving image $M$. A function $q^2$ is computed to estimate both the noise in local and global perspectives. Meanwhile, a certain number of neighborhood relations $N$ is also defined as to determine the kinds of self-similarities in the neighborhood. As extracting an image patch with a center at x, the measurement calculation of the self-similarity can be demonstrated as follows:

$$D(M, x, y) = \exp\left(\frac{S(x,y)}{q^2}\right) \quad x, y \in N \quad (1)$$

where *y* is defined as the center of another patch from one of the neighborhood *N*. This similarity metric helps to avoid the negative influence of image artifacts or random noise from the central patch extracted and prevent a direct adverse effect in calculation. Twelve distances between pairwise patches are calculated within six neighborhoods and concentrate in extracting the contextual neighboring information, instead of the direct shape representation.

## III. EXPERIMENTS

We evaluated the stability of our kidney atlas with a large cohort of multi-contrast unlabeled CT and a public portal venous contrast phase multi-organ labeled dataset. All CT data are unpaired and collected from different cohorts. With the use of both labeled and unlabeled datasets, we conducted comprehensive qualitative and quantitative evaluations for generalizing cross-modality information of kidney organs in CT.

*A. Datasets*

**Clinical Multi-Contrast Abdominal CT Cohort**: A large clinical cohort of multi-contrast CT was employed for abdominal and retroperitoneal organs registration. In total, 2000 patients' de-identified CT data were initially retrieved in de-identified form from ImageVU with the approval of Institutional Review Board (IRB). In these 2000 patients, since some had renal disease, criteria in ICD-9 codes and age range from 18-50 years old were set and applied to extract scans with healthy kidneys from all subjects. 720 subjects out of 2000 were identified after quality assessment and extract the corresponding contrast phase abdominal CT scans, which included 290 unlabeled CT volumes in total with: 1) non-contrast: 50 volumes, 2) early arterial: 30, 3) late arterial: 80 volumes, 4) portal venous: 100 volumes, 5) delayed: 30 volumes. All CT volumes are initially reoriented to standard orientation before further processing [13]. BPR was performed to each modality volumes and obtain the similar field of view with the atlas target. They were then resampled to the same resolution and dimensions with the atlas target for performing registration pipeline. We aim to adapt a generalized atlas framework for localizing the anatomical and contextual characteristics of kidney organs across multi-contrast.

**Multi-Organ Labeled Portal Venous Abdominal CT Cohort**: We used a separate healthy clinical cohort with 100 portal venous contrast phase abdominal CT volumes. The ground truth labels of 13 multiple organs are provided including: 1) spleen, 2) right kidney, 3) left kidney, 4) gall bladder, 5) esophagus, 6) liver, 7) stomach, 8) aorta, 9) inferior vena cava (IVC), 10) portal splenic vein (PSV), 11) pancreas, 12) right adrenal gland (RAD), 13) left adrenal gland (LAD), with which we conduct label transfer on this dataset to evaluate the generalizability and stability of the atlas template. In order to reduce the number of failed registrations to the atlas target, BPR is performed on abdominal and retroperitoneal regions only with soft-tissue window. To evaluate the atlas framework, the inverse transformation was applied to the multi-organ atlas label, and labels were transferred back to the spatial space of each portal venous phase CT.

*B. Evaluation Metrics*

We employed three commonly used metrics to evaluate the similarity between the prediction label from automatic models and the original ground truth label: 1) Dice score, 2) mean surface distance (MSD), and 3) Hausdorff distance (HD).

The definition of Dice is to measure the overlapping of volume between the segmentation label prediction and the ground truth segmentation label:

$$DICE(P, G) = \frac{2|P \cap G|}{|P| + |G|} \quad (2)$$

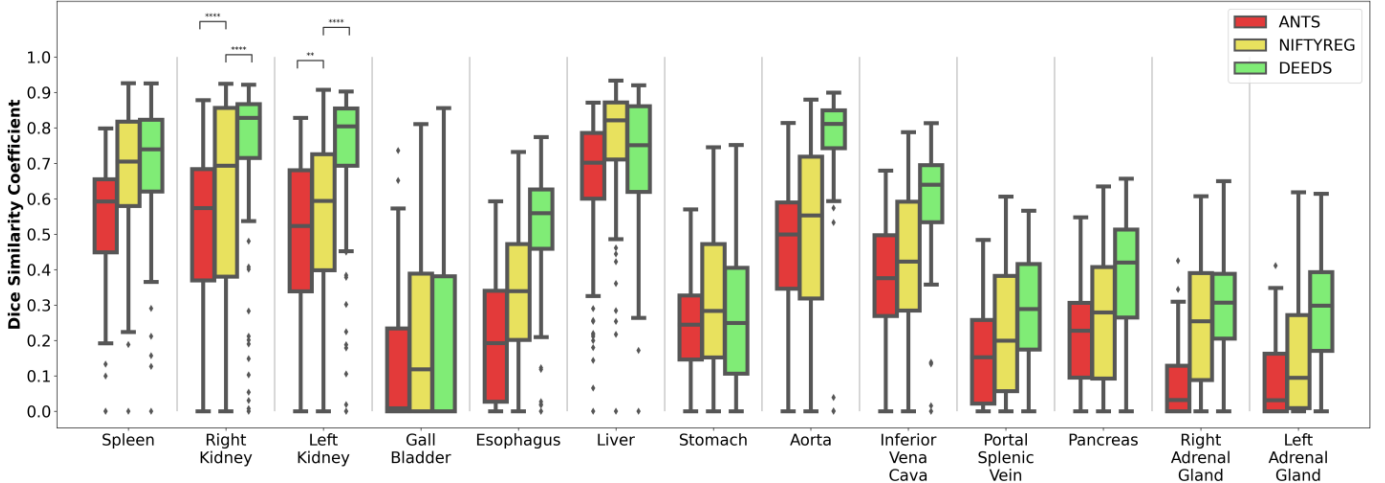

*Figure 3.* DEEDS registration outperforms the other two methods in an organ-wise manner. Significant increase of DICE is also demonstrated with median Dice score over 0.8 in the transferring result of both left and right kidneys using DEEDS (**: p < 0.01, ****: p < 0.0001, Wilcoxon signed-rank test).

where P is the predicted label from models and G is the original ground truth segmentation label, while ‖ is the L1 norm operation.

The rendered surface is another perspective which used to evaluate result. The 3-dimensional coordinates of vertices were initially extracted from both the prediction label and the ground truth label. The average distance and the Hausdorff distance between the sets of vertices coordinate were calculated as follows:

$$MSD(V_p, V_g) = avg \inf Dist(V_p, V_g) \quad (3)$$

$$HD(V_p, V_g) = sup \inf Dist(V_p, V_g) \quad (4)$$

where $V_p$ and $V_g$ represents the vertices coordinates of prediction label and ground truth label respectively, while *avg* refers to average, and *sup* and *inf* refers to the greatest lower bound and least upper bound of the distance function measure *Dist*.

## IV. RESULTS

Registrations were performed to the atlas target image with the pre-processed cropped volume of interest. We performed three registration methods to ensure the highest accuracy for the localization of both left and right kidneys: 1) ANTS [7, 14], 2) NiftyReg (NIFTYR) [7, 15], 3) DEEDS [7, 9, 10, 12]. The quantitative representation of each organ in the registered output was then demonstrated in terms of Dice score, MSD and HD, and illustrated the distribution of the performance across the multi-organ labeled portal venous cohort with Fig. 3 and Table 1.

As shown in Table I, DEEDS achieved the best overall mean Dice with the 2-stage registration pipeline across all 13 organs. From the demonstration of Fig.3, NiftyReg demonstrated a better performance in registering liver organ, while ANTs performed registration with inferiority across all organs comparing to the other two methods. As the atlas template is optimized for kidney organs, the Dice score of both left and right kidneys are separately computed to obtain the ability of kidneys localization with the atlas template.

TABLE I
EVALUATION METRIC ON 100 REGISTRATION FOR THREE REGISTRATION METHODS ON 13 ORGANS (MEAN ± STD)

| Methods | Dice Score | MSD (mm) | HD (mm) |
|---|---|---|---|
| ANTS (A) | 0.246 ± 0.224 | 12.8 ± 10.2 | 60.2 ± 47,5 |
| NIFTYR (A) | 0.270 ± 0.222 | 13.1 ± 11.1 | 62.1 ± 50.1 |
| DEEDS (A) | 0.200 ± 0.199 | 19.6 ± 18.2 | 80.9 ± 59.3 |
| ANTS | 0.319 ± 0.252 | 10.1 ± 9.00 | 51.6 ± 45.3 |
| NIFTYR | 0.406 ± 0.279 | 10.9 ± 13.8 | 55.0 ± 51.8 |
| DEEDS * | **0.496 ± 0.284** | **8.52 ± 17.1** | **41.6 ± 51.2** |

Note that p < 0.0001 with Wilcoxon signed-rank test **\***, A: affine registration only

DEEDS demonstrated the best performance in transferring the kidney information to the atlas space. NiftyReg and ANTs illustrated the lack of generalizability of transferring kidney organs and computed significant variance of Dice across all registrations. The reduction of variance and population of outliers are shown from boxplots with DEEDS, leading to a significant improvement of Dice score, MSD, and HD for both left and right kidneys comparing to the other two methods.

The Wilcoxon signed-rank test showed that DEEDS was significantly better (p < 0.0001) than all other methods in Dice [16]. Meanwhile, both left and right kidneys are registered to the atlas template with significant better result of using DEEDS (p < 0.0001). The median Dice of both transferred left and right kidney are above 0.8, while it is a significant boost comparing with the other two methods.

Apart from the quantitative result, the qualitative representations of registrations across multiple contrast phases are shown in terms of single subject registration, average template and variance mapping in Fig. 4. The multi-organ label provided general localization information of anatomical structures in atlas target image and compared with the deformed output from each contrast phase. The qualitative representation of the single subject registration demonstrated good localization ability of kidneys organ with the corresponding contrast characteristics, while the kidney organs in the early and late arterial phase subject is shown with small level of over-deformed. The average mapping of each contrast phase was then computed with all successfully registered contrast-corresponding volumes. Each average template is shown with stable abdominal body registration and a significant clear boundary of both left and right kidneys. The contrast characteristics of the kidneys were demonstrated and well allocated in the similar anatomical location of the atlas template. Apart from the contrast characteristics, the anatomy of kidney sub-structure and renal-related vessels can appear in the average template of early arterial, late arterial and delayed phase. To ensure the stability of transferring the kidneys' anatomical information, a variance map of each contrast phase template is also computed to demonstrate the voxel variability of each organ across the clinical cohort. The small variation in the kidney is illustrated with a color range from yellow to green, while significant variation in voxels is shown near the diaphragm region and the color range from orange to red indicated the highly deformed variability across the registered outputs. Overall, the anatomical and contrast characteristics of kidney organs can be both preserved and transferred to the atlas template with good stability.

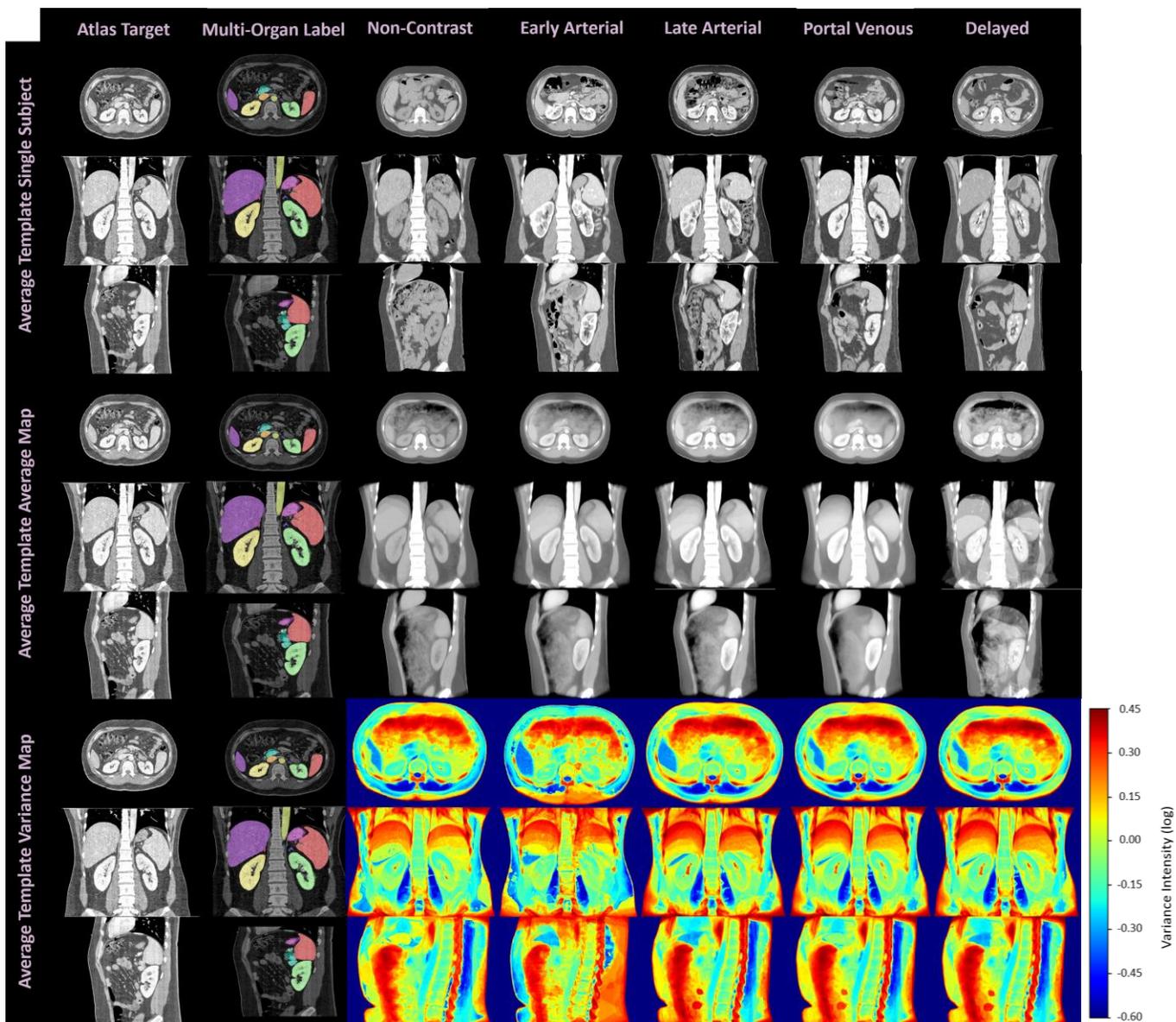

*Figure 4.* Qualitative representation of the single subject registrations, average mapping and variance mapping of contrast phases. The contrastive and morphological characteristics of kidney organs are demonstrated in the single subject registration and average mapping of each phases. Small variations are shown surrounding the kidney organs region in the variance mapping, while great variations are located in the diaphragm region nearby with liver and spleen.

## V. DISCUSSION

In this study, we constructed a healthy kidney atlas for five different contrast phases CT and generalized the kidney anatomical context across population demographics and the variation of contrast characteristics. High variance score is located near the diaphragm regions and the boundary of the abdominal body (see Fig. 4). Such variability is contributed to the large deformation of the lower and upper boundary of the volume interest and transfer specific organs' contextual information to other organs' anatomical locations. The low variance score in both the left and right kidney region indicated the stable registration of structural information, as the kidney organs are localized in the middle region of the volume interest with contrast. The surface rendering of the kidney across the morphological sizes visualizes the generalizability of the atlas template across shape variability (see Fig. 5). We adapt a 2D color-space checkerboard to visualize the deformation on the surface. The color of each box in the checkerboard pattern changes both horizontally and vertically. The color-boxes in atlas space are equivalent to that in the inverse original space. A stable deformation is demonstrated across small to large kidneys.

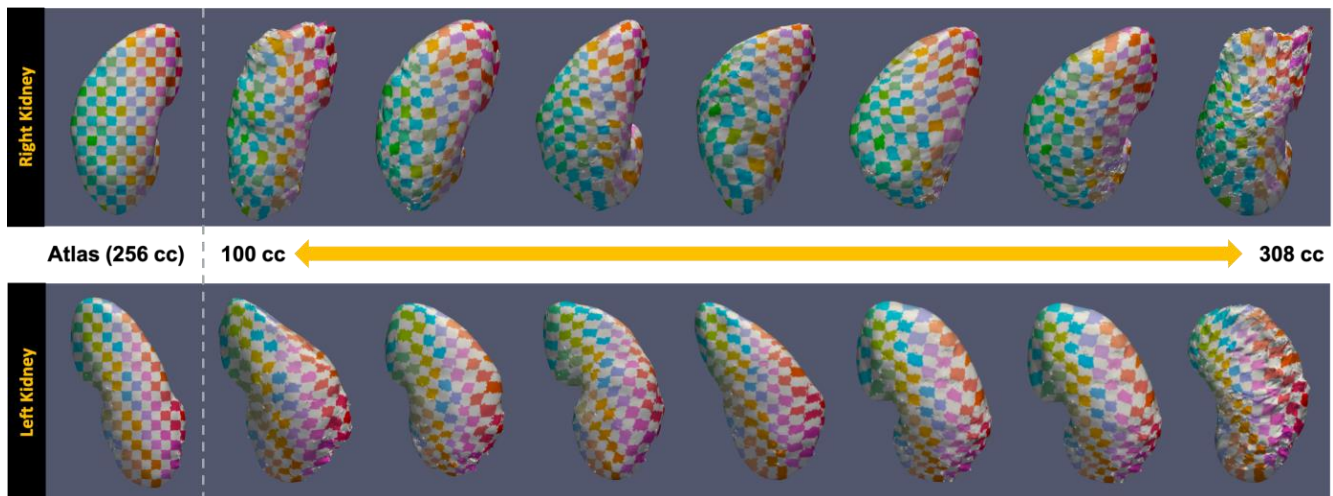

*Figure 5.* The surface rendering of the registered kidney with significant morphological variation are also illustrated. The 2D checkerboard pattern demonstrate the correspondence of the deformation from atlas space to the moving image space. A stable deformation across the change in volumetric morphology of kidney (100 cc to 308 cc) are demonstrated with the deformed checkerboard.

We choose the atlas template according to three conditions: 1) high-resolution (0.86 × 0.86 × 0.86 mm), 2) significant appearance of the morphological structure of kidneys, and 3) contrast characteristics of localizing kidneys. The high-resolution characteristics of the atlas template preserve highly detailed voxel-wise information across all organs. DEEDS provided the best overall performance with mean Dice of 0.50 [7]. However, the DEEDS performance cannot provide accurate measures towards the organ of interests due to high sensitivity of field of view, leading to significant deformation. Among the average template of each contrast phase in the presented figure, the portal venous phase average template provides a sharper and smoother abdominal body comparing to that of the other four phases. However, the portal venous phase average template is not suitable to be an unbiased template for registration. The small tissue voxels are difficult to represent and cannot provide sufficient accurate information for registration. Large population of clinical cohorts may need to be used to obtain a higher confidence level of voxel representation that is as sharp as the atlas target image.

Initially, one of the bottlenecks of providing a stable anatomical information transfer for organs is the registration method. The registration performance of kidneys may be affected by the secondary targets such as liver and spleen. Further optimization of the registration pipeline is needed for reducing the possibility of significant deformation. Instead of relying on similarity metric (mutual information, cross correlation, Hamming distances of the self-similarity context) as the loss function [17, 18], a learning-based method may be another promising direction to learn a registration function and predict the registration field for the moving images [19, 20]. However, most of the proposed learning-based pipeline is focused on the brain [21, 22]. The kidney atlas template provides an opportunity to further investigate learning-based approaches in abdominal and retroperitoneal regions.

The kidney atlas template also provides contributions in the segmentation of other abdominal and retroperitoneal organs. The high-quality atlas multi-organ label can be transferred with the inverse transformation and use to provide accurate measures for abdominal and retroperitoneal organs. Also, high quality labels can help perform training strategies to innovate automatic learning-based model. *Huo et al.* proposed a whole brain segmentation using spatially localized network tiles with the atlas-transferred label [23]. *Dong et al.* proposed left ventricle segmentation network, which integrate the ventricle atlas at echocardiogram into learning framework and provides consistency constraints with atlas label to perform accurate segmentations [24]. *Bai et al.* presented a population study of relating the phenome-wide association to the function of cardiac and aortic structures using machine-learning-based segmentation pipeline [25]. Clinical validation and phenotypic analysis can be performed and reveal biomarkers of specific organs in certain conditions such as disease pathogenesis, with high-quality segmentation labels. Further exploration can be investigated in the abdominal and retroperitoneal domain with the use of the high-quality atlas label.

## VI. Conclusion

This manuscript presents a healthy kidney atlas to generalize the contrastive and morphological characteristics across patients with significant variability in demographics and imaging protocols. Specifically, the healthy kidney atlas provides a stable reference standard for both left and right kidney organs in 3-dimensional space to transfer kidney information using an adapted registration pipeline. Significant variance on the field of view and the organ shape can be focused as the optimization parameters to reduce the possibility of failure registration. Potential future exploration with the use of the atlas template can be further investigated in both engineering and clinical perspectives, to provide better understandings and measures towards the kidneys.

ACKNOWLEDGEMENT

This research is supported by HuBMAP, NIH Common Fund and National Institute of Diabetes, Digestive and Kidney Diseases U54DK120058 (Spraggins), NSF CAREER 1452485, NIH 2R01EB006136, NIH 1R01EB017230 (Landman), and NIH R01NS09529. This study was in part using the resources of the Advanced Computing Center for Research and Education (ACCRE) at Vanderbilt University, Nashville, TN. The identified datasets used for the analysis described were obtained from the Research Derivative (RD), database of clinical and related data. ImageVU and RD are supported by the VICTR CTSA award (ULTR000445 from NCATS/NIH) and Vanderbilt University Medical Center institutional funding. ImageVU pilot work was also funded by PCORI (contract CDRN-1306-04869).